\documentclass{aa}
\usepackage{graphicx}
\usepackage{natbib}

\def\dpi#1{$^2\Pi_{{#1}/2}$}
\def\micron{$\mu$}

\begin{document}
\newcommand{\mc}{\multicolumn {2}{c}}

  \title{A high-sensitivity OH 5-cm line survey in late-type
  stars\footnote{Table \ref{table1} is also available in electronic
  form at the CDS via anonymous ftp to cdsarc.u-strasbg.fr
  (130.79.128.5) or via
  http://cdsweb.u-strasbg.fr/cgi-bin/qcat?J/A+A/}}

  \subtitle{}

  \author{J.-F. Desmurs\inst{1}
     \and A. Baudry\inst{2}
     \and P. Sivagnanam\inst{3}
     \and C. Henkel\inst{4}
         }


    \offprints{J.-F. Desmurs}

    \institute{Observatorio Astron\'omico Nacional (IGN), Apartado 1143,
                 E--28800 Alcal\'a de Henares, Spain \\
                 email: desmurs@oan.es
          \and
              Observatoire de l'Universit\'e  de Bordeaux, URA 352
              du CNRS, BP89, F-33270 Floirac, France.
          \and
              Observatoire de Paris, 5, Place Jules Janssen
              F-92195 Meudon CEDEX, France.
          \and
              Max-Planck-Institut f\"ur Radioastronomie,
              Auf dem H\"ugel 69, D-53121 Bonn, Germany.
              }

    \date{Receive7 May 2002; Accepted 27 August 2002}
    \titlerunning {Survey 6GHz OH Maser in OH Stars.}

    \abstract{ We have undertaken a comprehensive search for 5-cm
excited OH maser emission from evolved stars representative of various
stages of late stellar evolution.  Observed sources were selected from
known 18-cm OH sources.  This survey was conducted with the 100-m
Effelsberg telescope to achieve high signal to noise ratio observations
and a sensitivity limit of about 0.05 to 0.1 Jy. A total of 64 stellar
sources were searched for both main line and satellite line emission.
We confirm the previous detection of 5 cm OH in Vy 2-2, do not confirm
emission from NML-Cyg and do not report any other new detection within
the above sensitivity limit.

Implications of these results on the pumping mechanism of the OH
radical in circumstellar envelopes are briefly discussed.
             
           }
    \maketitle
\section{Introduction}
\label{intro}

\begin{figure*}
\vspace{15cm}
\includegraphics{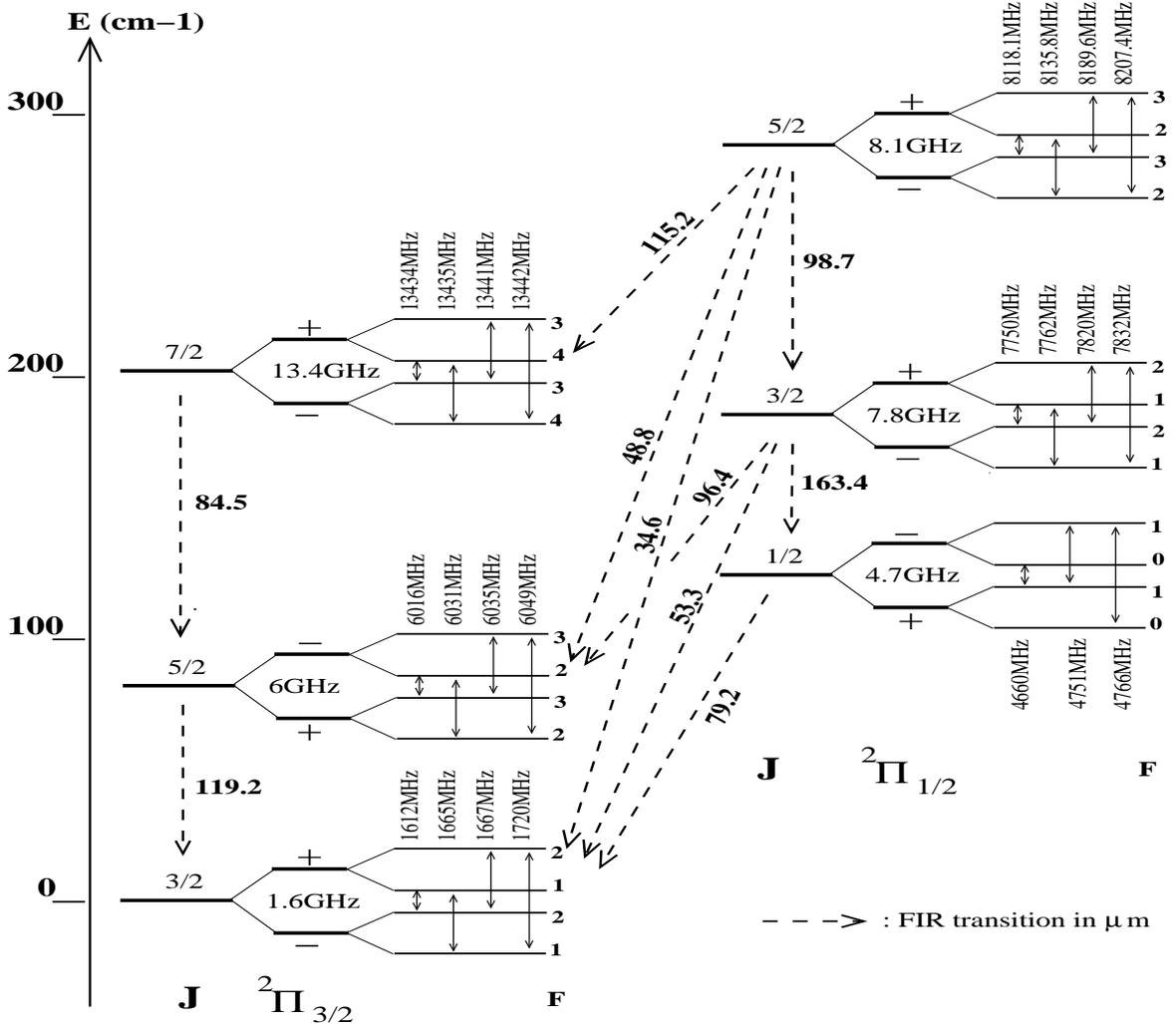}
	\caption[]{The energy level diagram for the $^{ 2}{ \Pi }_{
3/2}$ and $^{ 2}{ \Pi }_{ 1/2}$ ladders of OH. $\Lambda$ doubling (not to
scale) and parities are shown in each case. Transitions between the
$F=3$ and 2 hyperfine levels, for $^{ 2}{ \Pi }_{3/2} , J=5/2$, give
rise to the four 6 GHz lines} 

\label{OH_energy}
\end{figure*}

Emission from the first two rotationally excited states of OH was first
discovered by \cite{zuc68} and \cite{yen69} for the $^{ 2}{ \Pi }_{
1/2}{ }, J=1/2$ and $^{ 2}{ \Pi }_{ 3/2}, J=5/2$ states, respectively.
The $^{ 2}{ \Pi }_{3/2} , J=5/2$ state of OH lies immediately above the
ground-state and gives rise to four hyperfine transitions, with the
$F=3-3$ and $2-2$ main lines at 6035.092 and 6030.747 MHz and the
$F=3-2$ and $2-3$ satellite lines at 6049.084 and 6016.746 MHz,
respectively (Fig. \ref{OH_energy}).
The theoretical treatment of OH excitation in star-forming regions has
progressed significantly in recent years \citep[see][]{ces91,gra92,
pav96}, and good predictions of relative OH line intensities
can be made on the basis of these models,
which show the importance of multi-line studies.  In the circumstellar
environment of late-type stars the model developed by \cite{eli76}
successfully explains the excitation of strong 1612 MHz emission. This
results from a cascade of the OH population down to the J=1/2 and 3/2
states after far infrared photons at 34.6 micron and 53.3 micron  (see Figure
\ref{OH_energy}) have excited the OH to the $^{ 2}{ \Pi }_{ 1/2}{ },
J=5/2$ and $3/2$ states. There are enough far infrared photons to
excite the 1612 MHz line \citep[e.g.][]{epc80}.  However, it is only
recently that the direct detection of the 34.6 microns absorption line
has been reported with the ISO telescope toward IRC+10420 
\citep[see][]{syl97}. Besides the conspicuous 1612 MHz line emission, 18 cm
main line emission is often observed in late-type stars. Conditions for
this emission are carefully investigated in the work of
 \cite{col94, col95} and we discuss later in
this work the implication of their excitation mechanism for the J=5/2
state of OH.

The main goal of the present observations was to survey the 5cm
$\Lambda$ doublet lines of OH in a number of stars ranging from
typical Miras to OH/IR objects or pre-planetary nebulae.  These stars
sample various late stages of stellar evolution.  In addition,
observations of the $^{ 2}{ \Pi }_{3/2} , J=5/2$ state lying
immediately above the ground-state provide a critical test for OH
excitation models.

\par

These observations and our results obtained in 64 sources are presented
in Sects. 2 and 3. Some OH properties of selected stars are also
presented in Sect. 3. In Sect. 4 we discuss stellar OH pumping schemes
and variability of OH main line emission sources in the light of our
results.

\section{Observations}
Observations were made with the 100-m antenna at Effelsberg\footnote
{The 100-m telescope at Effelsberg is operated by the
Max-Planck-Institut f{\"u}r Radioastronomie (MPIfR) on behalf of the
Max-Planck-Gesellschaft (MPG).} in December 1999. We used a cooled HEMT
dual-channel receiver connected to only one sense of polarization, Left
Circular Polarization ($LCP$).  The system temperature was $\approx$ 60
K ($T_{\rm mb}$) including ground pick up and sky noise.  We used a
position switching observing mode with the reference position offset by
600'' in longitude from the source position (the half-power beamwidth
of the telescope at 6 GHz is 130''). The new auto-correlator AK90 was
split into 2 bands of 20 MHz each thus allowing us to simultaneously
observe the 2 main lines and the two satellites lines of the J=5/2
state.  There were 4096 channels per band giving a channel separation
of 4.9 kHz and thus an effective spectral velocity resolution of 0.29
km s$^{-1}$.  Proper functioning of the system was checked by
observations of the strong 5cm OH emission from the two compact HII
regions W3(OH) and ON1 (see Table \ref{table1}).

Calibration of the data followed the procedure used in the 6 GHz survey
of star-forming regions made by \cite{bau97}. OH spectra were
calibrated in terms of the noise source coupled to one polarization
channel and the flux density scale was determined by observations of
NGC 7027 \citep{ott94}.
The noise tube was calibrated in Jy assuming that the 6 GHz flux
density of NGC7027 was 5.9 Jy. We estimate that the flux density scale
uncertainty is within 10\%.
All spectra were calibrated in terms of single
polarization flux densities. This is one-half of the two polarization
flux density. For possible 5 cm radio interference, we proceeded
as in \cite{bau97}.

Our input catalog is listed in Table \ref{table1}. It is based on 18 cm
OH data. We selected sources which clearly exhibit the 1612 MHz
satellite line and/or the 1665/1667 main lines. By these means we
obtained targets with noticeable amounts of OH molecules and IR photons,
that are not excessively distant in order to be detected.

\begin{figure*}[pt]
\vspace{8cm}
\includegraphics{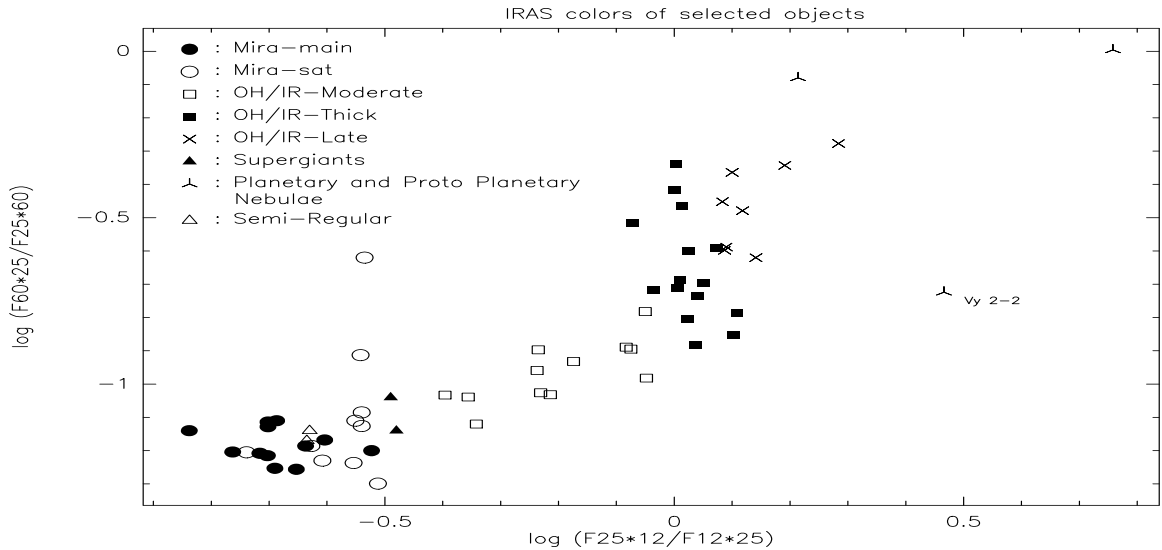}
	\caption[]{Uncorrected IRAS colors from IRAS flux measurements
         available of the sources in our sample. (NML Cyg is not
         included as there is no IRAS measurement available).}
\label{iras_flux}
\end{figure*}

The sources are essentially OH Miras with thin or moderately thick
envelopes, and thick OH/IR objects. Bright Miras with both satellite
and main lines were selected, from the \cite{siva88} comprehensive OH
survey of the 1-kpc solar neighborhood. Most of them are also known as
22 GHz water maser sources. From the \cite{dav93} Nan\c{c}ay survey of
OH/IR objects, we retained bright 1612 MHz sources with detected 1667
MHz emission. 
Targets have been chosen in order to sample objects lying
along the sequence drawn by evolved stars in the IRAS two-color diagram
(see Fig.  \ref{iras_flux}). They mostly belong to the AGB. This
sequence of envelope thicknesses is thought to reflect various
evolutionary stages, and a certain initial mass distribution. Given 
our selection criteria we favored the best candidate stars for showing
OH maser emission i.e. the nearby OH/IR stars with high luminosity and
large amounts of dust and infrared photons.

We also included in our survey a few typical Semi-Regular variables,
Proto-Planetary Nebulae (PPN), and Planetary Nebulae (PN), that are
thought to be linked to the same evolutionary tracks. We added some
red Supergiants (SG). The latter are luminous massive stars that are not
on the AGB, but which usually present somewhat similar circumstellar
envelopes.  In our input catalog are NML Cyg (SG) and Vy\,2-2 (PN),
for which tentative detections at 5 cm have been reported in the
literature.

\begin{figure*}[pt]
\vspace{7cm}
\includegraphics{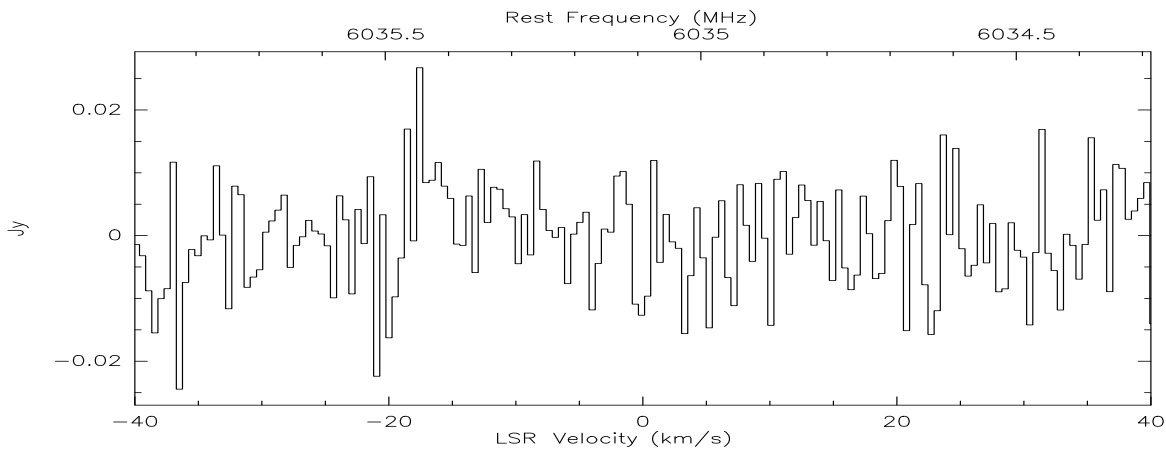}
	\caption[]{The 6 GHz spectrum obtained in Dec 1999 for NML Cyg,
        the rms at 1$\sigma$ is $\sim$6mJy (LCP spectrum). No detection
	reported.}
\label{spectra_nml}
\end{figure*}

\begin{figure*}[pt]
\vspace{7cm}
\includegraphics{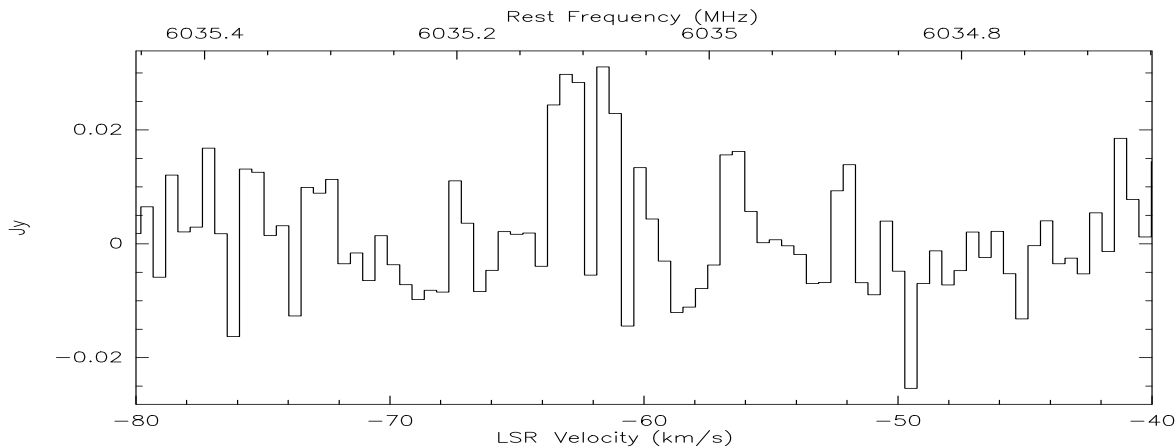}
	\caption[]{ 6035 MHz OH spectrum obtained in Dec 1999 from Vy
        2-2. The line intensity is in Jy for single polarization,
        the rms at 1$\sigma$ is $\sim$6mJy.}
\label{spectra_vy22}
\end{figure*}

\begin{figure*}[pt]
\vspace{7cm}
\includegraphics{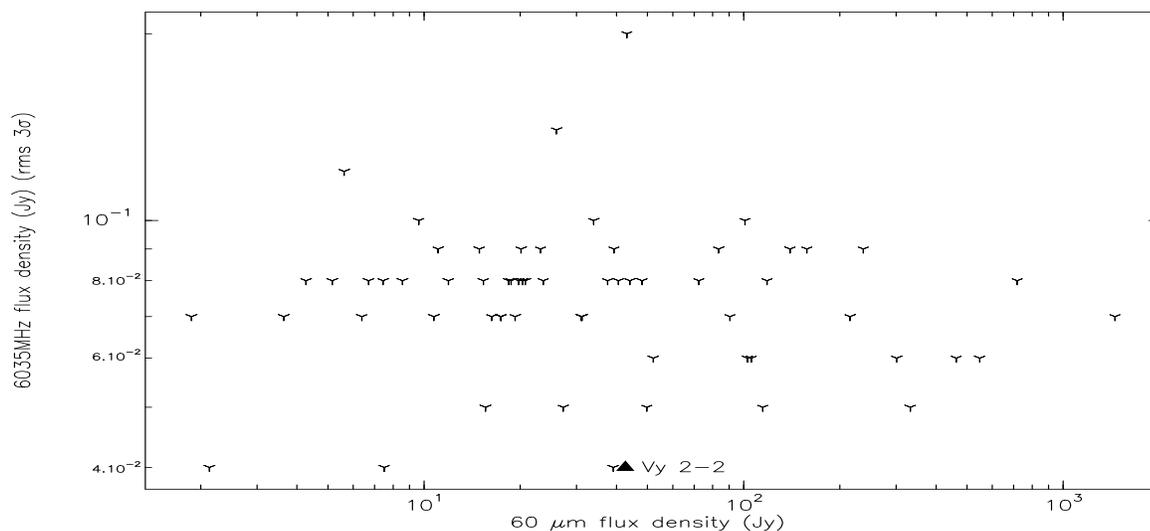}
	\caption[]{IRAS 60 \micron m versus the OH 6035 MHz 3$\sigma$
         flux density limit (channel 0.25 km/s). The filled triangle 
         corresponds to the only detection in our sample, Vy2-2,
         the detected flux is given.}
\label{flux_iras}
\end{figure*}

%
\oddsidemargin -1cm
\setcounter {table} 0
\addtocounter{footnote}{-2}
\begin{table*}[t]
 \begin{center}
 \caption{Observations of the 6GHz OH maser transitions in AGB stars}
 \label{table1}
 \small
 \oddsidemargin -2cm
 \begin{tabular}{ c c c c c c c }
 \hline
 \bf{IRAS Source}&\bf{Other Name}&\bf{Type\footnotemark}&\mc
{\bf{Observed Coordinates}}&\bf{Observed }&{\bf{Sensitivity
\footnotemark (Jy)}}\\ \bf{ }&\bf{ }&\bf{ }&\mc{\bf{J2000.0
\footnotemark}}&\bf{Velocity}&\bf{ (at 3$\sigma$) }\\
 \bf{ }&\bf{ }&\bf{ }&\bf{R.A.}&\bf{Dec.}&\bf{Range, LSR}&\bf{6GHz}\\
 \bf{ }&\bf{ }&\bf{ }&\bf{h m s}&\bf{${ }^0$ '
 ''}&\bf{(km.s$^{-1}$)}&\bf{LCP}\\
 \hline
 \rm
$02232+6138$ &W3(OH)&UCHII& 2:27:03.866 & $+$61:52:24.82 & -939/+998,($-$45.0)& E \\
$20081+3122$ &ON1&UCHII& 20:10:09.073 & $+$31:31:34.40 & -894/1043,($+$00.0)& E \\
 \hline
$00007+5524$ &Y Cas& Mira main& 00:03:21.605 & $+$55:40:48.21 & -911/1026,($-$17.0)&  0.08   \\
$01037+1219$ &IRC+10011& OH/IR Moderate& 01:06:25.965 & $+$12:35:53.07 & -886/1051,($+$08.0)&  0.07   \\
$02168-0312$ &OMI Cet& Mira main& 02:19:20.679 & $-$02:58:39.21 & -848/1089,($+$46.0)&  0.06   \\
$02192+5821$ &S Per& SG & 02:22:51.689 & $+$58:35:11.76 & -932/1005,($-$38.0)&  0.08   \\
$02251+5102$ &RR Per& Mira main& 02:28:29.318 & $+$51:16:17.60 & -894/1043,($+$00.0)&  0.08   \\
$02420+1206$ &RU Ari& Mira sat& 02:44:45.184 & $+$12:19:08.15 & -874/1063,($+$20.0)&  0.07   \\
$03206+6521$ &OH 138.0+7.2&OH/IR Moderate& 03:25:08.552 & $+$65:32:05.42 & -931/1006,($-$37.5)&  0.08   \\
$03293+6010$ &OH 141.0+3.5&OH/IR Moderate& 03:33:30.528 & $+$60:20:09.10 & -951/986,($-$57.2)&  0.07   \\
$03507+1115$ &IK Tau& Mira sat& 03:53:28.983 & $+$11:24:20.02 & -861/1076,($+$33.0)&  0.05   \\
$05073+5248$ &NV Aur&OH/IR Moderate& 05:11:19.747 & $+$52:52:27.79 & -891/1046,($+$02.9)&  0.08   \\
$05131+4530$ &RAFGL 712&OH/IR Moderate& 05:16:47.103 & $+$45:34:03.76 & -926/1011,($-$32.3)&  0.09   \\
$05528+2010$ &U\ Ori&Mira sat& 05:55:49.264 & $+$20:10:30.77 & -933/1004,($-$39.5)&  0.04   \\
$05559+7430$ &V Cam& Mira main& 06:02:32.779 & $+$74:30:27.26 & -887/1050,($+$06.5)&  0.07   \\
$06297+4045$ &OH 174.7+13.5&OH/IR Moderate& 06:33:14.921 & $+$40:42:49.71 & -910/1027,($-$16.3)&  0.08   \\
$06363+5954$ &U Lyn& Mira main& 06:40:46.320 & $+$59:52:00.23 & -904/1033,($-$10.0)&  0.07   \\
$06500+0829$ &GX Mon& Mira sat & 06:52:42.408 & $+$08:25:23.33 & -905/1032,($-$11.0)&  0.06   \\
$07209-2540$ &VY\ Cma&SG& 07:22:58.17 & $-$25:46:02.95 & -872/1065,($+$22.0)&  0.07   \\
$07399-1435$ &OH231.8+4.2& PPN & 07:42:16.738 & $-$14:42:14.04 & -878/1059,($+$16.0)&  0.06   \\ 
$07422+3054$ &AU Gem& Mira Main& 07:45:27.590 & $+$30:46:43.11 & -890/1047,($+$04.0)&  0.04   \\
$07585-1242$ &U Pup& Mira sat & 08:00:50.596 & $-$12:50:31.88 & -912/1025,($-$18.0)&  0.04   \\
$08357-1013$ &OH235.3+18.1& OH/IR Moderate & 08:38:08.805 & $-$10:24:16.95 & -894/1043,($+$00.0)&  0.07 \\
$09425+3444$ &R Lmi& Mira main& 09:45:34.314 & $+$34:30:42.91 & -894/1043,($+$00.0)&  0.14   \\
$09448+1139$ &R Leo& Mira main & 09:47:33.446 & $+$11:25:43.52 &-893/1044,($+$01.0)&  0.05   \\
$10580-1803$ &R Crt& SR & 11:00:33.822 & $-$18:19:29.54 & -884/1053,($+$10.0)&  0.05   \\
$12562+2324$ &T Com& Mira sat & 12:58:38.600 & $+$23:08:23.03 & -879/1058,($+$15.0)&  0.07   \\
$13001+0527$ &RT Vir& SR & 13:02:37.908 & $+$05:11:08.56 & -876/1061,($+$18.0)&  0.09   \\
$14247+0454$ &RS Vir &Mira sat & 14:27:16.393 & $+$04:40:40.38 & -894/1043,($+$00.0)&  0.08   \\
$15193+3132$ &S Crb& Mira main& 15:21:24.366 & $+$31:22:02.05 & -894/1043,($+$00.0)&  0.07   \\
$15255+1944$ &WX Ser& Mira sat & 15:27:47.009 & $+$19:34:02.33 & -888/1049,($+$06.0)&  0.08   \\
$16235+1900$ &U Her&Mira main& 16:25:47.411 & $+$18:53:32.98  & -907/1030,($-$13.0)&  0.05   \\
$18006-1734$ &GLMP 704&OH/IR Late& 18:03:36.581 & $-$17:34:00.59 & -870/1067,($+$23.9)&  0.10   \\
$18071-1727$ &OH 12.8 +.9&OH/IR Late& 18:10:05.813 & $-$17:26:35.22 & -868/1069,($+$25.9)&  0.09   \\
$18081-0338$ &&OH/IR Thick& 18:10:49.074 & $-$03:38:14.46 & -889/1049,($+$05.6)&  0.08   \\
$18100-1915$ &GLMP 740&OH/IR Thick& 18:13:02.725 & $-$19:14:20.42 & -876/1061,($+$17.6)&  0.09   \\
$18107-0710$ &&OH/IR Thick& 18:13:29.720 & $-$07:09:48.92 & -876/1061,($+$17.9)&  0.08   \\
$18135-1456$ &OH 15.7 +.8&OH/IR Late& 18:16:26.004 & $-$14:55:13.43 & -895/1042,($-$01.1)&  0.09   \\
\hline
\end{tabular}
\end{center}
\end{table*}

\setcounter {table} 0
\oddsidemargin -1cm
\begin{table*}[t]
 \begin{center}
 \caption{(continue) Observations of the 6GHz OH maser transitions in AGB stars}
 \small
 \oddsidemargin -2cm
\begin{minipage}{\textwidth}
\renewcommand{\thempfootnote}{\arabic{mpfootnote}}
 \begin{tabular}{ c c c c c c c }
 \hline
 \bf{IRAS Source}&\bf{Other Name}&\bf{Type \footnote{UCHII: Ultra
 Compact HII region, SG: Supergiante, SR: Semi Regular, PN: Planetary
 Nebulae, PPN: Proto Planetary Nebulae.}}&\mc{\bf{Observed Coordinates}}
 &\bf{Observed}& {\bf{Sensitivity \footnote{E = Emission; the upper
 limits correspond to 3$\sigma$.} (Jy)} }\\ \bf{ }&\bf{ }&\bf{
 }&\mc{\bf{J2000.0}\footnote {B1950 coordinates processed to J2000
 within GILDAS package.}}&\bf{Velocity}&\bf{ (at 3$\sigma$) } \\
 \bf{ }&\bf{ }&\bf{ }&\bf{R.A.}&\bf{Dec.}&\bf{Range, LSR}&\bf{6GHz}\\
 \bf{ }&\bf{ }&\bf{ }&\bf{h m s}&\bf{${ }^0$ '
 ''}&\bf{(km.s$^{-1}$)}&\bf{LCP}\\
 \hline
 \rm
$18152-0919$ &OH 20.8 +3.1&OH/IR Thick& 18:17:58.662 & $-$09:18:42.44  & -867/1070,($+$26.7)&  0.08   \\
$18198-1249$ &OH 18.3 +.4&OH/IR Late& 18:22:43.046 & $-$12:47:40.94 & -846/1091,($+$48.0)&  0.10   \\
$18262-0735$ &&OH/IR Thick& 18:28:59.446 & $-$07:33:25.44 & -815/1122,($+$78.9)&  0.08   \\
$18268-1117$ &OH 20.4 -.3&OH/IR Thick& 18:29:35.755 & $-$11:15:53.97 & -852/1085,($+$41.8)&  0.08   \\
$18266-1239$ &V435 Sct&OH/IR Thick& 18:29:28.605 & $-$12:37:40.55 & -844/1093,($+$50.0)&  0.08   \\
$18348-0526$ &V437 Sct&OH/IR Thick& 18:37:31.986 & $-$05:23:59.35 & -867/1070,($+$27.2)&  0.06   \\
$18432-0149$ &V1360 Aql&OH/IR Thick& 18:45:52.691 & $-$01:46:43.27 & -828/1109,($+$66.2)&  0.08   \\
$18460-0254$ &V1362 Aql&OH/IR Late& 18:48:40.999 & $-$02:50:22.29 & -795/1142,($+$99.0)&  0.09   \\
$18488-0107$ &V1363 Aql&OH/IR Late& 18:51:25.772 & $-$01:03:54.48 & -818/1119,($+$75.8)&  0.08   \\
$18525+0210$ &&OH/IR Thick& 18:55:04.815 & $+$02:14:41.21 & -823/1113,($+$70.2)&  0.07   \\
$18549+0208$ &OH 35.6 -.3&OH/IR Thick& 18:57:26.573 & $+$02:12:11.24 & -816/1121,($+$77.9)&  0.09   \\
$18560+0638$ &V1366 Aql&OH/IR Moderate& 18:58:30.142 & $+$06:42:55.91 & -874/1063,($+$19.7)&  0.10   \\
$19039+0809$ &R Aql& Mira sat & 19:06:22.196 & $+$08:13:48.16 & -846/1091,($+$48.0)&  0.09   \\
$19065+0832$ &OH 42.6 +.0&OH/IR Late& 19:08:58.368 & $+$08:37:47.08 & -841/1096,($+$53.1)&  0.07   \\
$19071+0946$ &OH 43.8 +.5&OH/IR Thick& 19:09:31.065 & $+$09:51:54.41 & -885/1052,($+$09.1)&  0.12   \\
$19161+2343$ &&OH/IR Moderate& 19:18:14.537 & $+$23:49:26.23 & -865/1072,($+$28.6)&  0.07   \\
$19219+0947$ &Vy\ 2-2&PN    & 19:24:22.078 & $+$09:53:55.82 & -956/981,($-$44.3)&  E/0.02   \\
$19244+1115$ &V1302 Aql&OH/IR Moderate& 19:26:47.588 & $+$11:21:14.77 & -849/1088,($+$45.0)&  0.08   \\
$19343+2926$ &M1-92& PPN & 19:36:16.768& $+$29:32:15.80& -894/1043,($+$0.00)& 0.24  \\
$19352+2030$ &&OH/IR Thick& 19:37:23.395 & $+$20:39:21.86 & -889/1048,($+$05.0)&  0.20   \\
$20043+2653$ &GLMP 972&OH/IR Thick& 20:06:22.890 & $+$27:02:11.23 & -898/1039,($-$04.6)&  0.08   \\
$20047+1248$ &SY Aql& Mira sat & 20:07:05.694 & $+$12:57:07.39 & -943/994,($-$49.0)&  0.08   \\
NML Cyg  &&SG & 20:46:25.941 & $+$40:06:56.09 & -894/1043,($+$00.0)&  0.02   \\
$20491+4236$ &OH 83.4 -.9&OH/IR Moderate& 20:50:57.766 & $+$42:48:04.31 & -932/1005,($-$38.4)&  0.05   \\
$21554+6204$ &GLMP 1048&OH/IR Thick& 21:56:58.184 & $+$62:18:43.62 & -914/1022,($-$20.6)&  0.06   \\
$22177+5936$ &NSV 25875&OH/IR Moderate& 22:19:27.806 & $+$59:51:21.74 & -919/1017,($-$25.6)&  0.07   \\
$23041+1016$ &R Peg &Mira main& 23:06:38.829  & $+$10:32:37.94 & -870/1067,($+$24.0)&  0.09   \\
$23558+5106$ &R\ Cas&Mira main& 23:58:24.683 & $+$51:23:18.18 & -870/1067,($+$24.0)&  0.06   \\
\hline
\end{tabular}
\end{minipage}

\end{center}
\end{table*}

\section{Results}

\subsection{Previous Surveys}
Previous attempts made to search for excited OH from circumstellar
envelopes gave only negative or controversial results with the
exception of one object. As far as we are aware only a few
searches for J=1/2 and 5/2 OH emission at 4.7 and 6 GHz from stars have
been undertaken \citep[see][]{tha70, zuc72, bau74, cla81, jew85}.  The
latter work was the most sensitive search for excited OH from stars yet
performed.  \cite{zuc72} reported weak 6035 MHz ($^{ 2}{ \Pi }_{3/2} ,
J=5/2$) emission from NML Cyg and \cite{cla81} reported weak 4751 MHz
($^{ 2}{ \Pi }_{1/2} , J=1/2$) emission from AU Gem.  However, both
detections were not confirmed by \cite{jew85}. On the other hand,
Jewell et al., reported weak 6035 MHz maser emission from the planetary
nebula Vy 2-2 appearing at the same velocity, --62km\,s$^{-1}$, as the
peak 1612 MHz maser emission detected by \cite{dav79}.

\subsection {New Effelsberg survey}

In Table \ref{table1}, we list the 64 late type stars observed by
us. For all sources, the velocity range of search for emission is given
(with the systemic velocity in parenthesis), together with the
sensitivity limit achieved in our new survey at 3$\sigma$.  The average
noise level reached in our survey is (at 3$\sigma$ with a channel width
of 0.29 km\,s$^{-1}$) around 80 mJy ; in comparison, \cite{jew85} have
reached about 230 mJy (with a channel width of 0.06 km\,s$^{-1}$).

Of the 64 sources observed, no one exhibits a clear emission or
absorption signal. There are however two sources with tentative
detections, NML Cyg (see Fig. \ref{spectra_nml}) and Vy 2-2
(Fig. \ref{spectra_vy22}). For NML Cyg, we reached a sensitivity of
20 mJy (at 3$\sigma$ level) over the observed LSR velocity range. The
0.8 K (2.2Jy) signal reported by \cite{zuc72} and lying close to
+5\,km\,s$^{-1}$, would have been easily detected by us. However, we
can not exclude that the emission varies with time. The tentative
feature at about --17\,km\,s$^{-1}$ (Fig. \ref{spectra_vy22}) is only
detected at a $\sim$3$\sigma$ level and is therefore not convincing,
but we note that 1612\,MHz line emission at --18\,km\,s$^{-1}$ has been
reported previously \citep[see e.g. ][]{eng79}.

The case of Vy 2-2 is different.  With an integrated intensity of
$\sim$48 mJy km\,s$^{-1}$, we have obtained a 6$\sigma$ detection. 
Only the F=3--3 maser line transition lying at 6035 MHz was
detected. No absorption or emission can be observed for the other
transitions. Fig. \ref{spectra_vy22} shows the observed 6035 MHz
spectrum. The parameters and uncertainties (1$\sigma$) of Gaussian fits
to the detected features are displayed in Table \ref{table_vy2-2}.  The
derived apparent luminosity is 1.1 Jy\,km\,s$^{-1}$kpc$^2$
\citep[assuming a distance of 3.8kpc,][]{ben01}.  The lack of F=2-2
emission\footnote{In the LTE approximation we obtain F=3-3/F=2-2 approx
1.4 whereas we observe here $>$2.} and the narrow F=3-3 linewidth
suggest that the observed F=3-3 line results from a maser
process. However, only interferometric observation could give a
definitive proof of it.

\setcounter {table} 1
\begin{table}[t]
\caption{Gaussian line parameters of the 6035 MHz OH emission line of
Vy 2-2}
\label{table_vy2-2}
\begin{center}
\small
\oddsidemargin -2cm
\begin{tabular}{|c|c|c|}
\hline
\bf{Velocity}&\bf{Peak flux density }&\bf{Linewidth}\\
\bf{ (kms$^{-1}$) }&\bf{ (mJy) }&\bf{(kms$^{-1}$)  }\\
\hline
 -63.0$\pm$0.14 &  38$\pm$8 & 1.14$\pm$0.17 \\
 -61.6$\pm$0.14 &  39$\pm$8 & 0.84$\pm$0.17 \\
\hline
\end{tabular}
\end{center}
\end{table}

This detection is consistent with the results of \cite{jew85} who
observed maser emission at nearly the same velocity ($\sim$
--62.3\,km\,s$^{-1}$) and with about the same line width
($\sim$1.5\,km\,s$^{-1}$) but with a peak flux intensity four times
stronger (0.15 Jy). The presence of the two features
(Fig.\ref{spectra_vy22}) is likely real.  After splitting our data in
two equal parts, the same two components appear.  In another data
reduction test, we have degraded our spectral resolution. This yields
one single feature with a line width of $\sim$2.5 km\,s$^{-1}$,
i.e. twice the line width observed by Jewell et al., centered around
-62.3 km\,s$^{-1}$. Our observations and data reduction confirm long
term OH emission from Vy2-2.

\subsection{Vy\,2-2}

As is the case for other Galactic planetary nebulae, the distance to Vy
2-2 (G045.4-02.7) is poorly known.  Previous attempts to determine the
distance have resulted in a wide range of estimates. Those estimates
put this object from 1.9 kpc \citep[see][]{ack78} based on an optical
calibration to a kinematic distance of 20 kpc \citep{dav79}.  The most
recent estimate, based on a compilation of previous measurements
\citep[see][]{ben01} gives a distance of 3.8 kpc.
Vy 2-2 is a source of free-free radio continuum radiation and
dust-type infrared emission. VLA maps show a slightly elongated
continuum source \citep{sea83}. The continuum emission originates from
a compact (diameter$\sim$0.5'') and narrow (thickness $\sim <$ 0.''12)
shell of ionized gas. This ionized region is surrounded by an extended
halo of over 25'' in radius, detected through its H$\alpha$ line
emission \citep[see][]{mir91}.
From the visibility analysis, \cite{chr98} estimate an angular
expansion of 1.13$\pm$0.12 mas/yr$^{-1}$. This would give for a
distance of 3.8 kpc an expansion velocity of about 20km/s, in
contradiction with the expansion velocity of 6 km\,s$^{-1}$ measured by
\cite{mir91} in the equatorial plane and qualified to be slow.  Taking
a systemic velocity for the source of --44.3$\pm$1.0km\,s$^{-1}$
\citep[tentative detection of][]{kna85} this would give a blue-shifted
velocity for the OH maser of about 20 km\,s$^{-1}$. The inferred
expansion velocity is then in good agreement with
the value derived by \cite{chr98} for a distance of 3.8 kpc.  The
kinematic age of the nebula they derived is 213 years and
supports the conclusion that this object is a very young planetary
nebula.
The temperature of the central star is estimated to be greater than
35000K \citep[see][]{zij89,cle89}. The dust color temperature was
estimated by \cite{coh74} to be less than 190K. 

\cite{jew85} and \cite{coh91} searched without success for
$^{2}{\Pi}_{1/2} , J=1/2$ maser emission (down to a $3\sigma$ limit of
$\sim$0.25Jy). The 1612 MHz maser emission, the only ground-state
maser transition observed, was first detected by \cite{dav79}.
\cite{sea83} located the maser at the front edge of the ionized shell,
coincident with a shock front and an ionization front, placing the OH
maser on the near side of the expanding shell and thus providing an
explanation for the blue-shifted maser feature. This is consistent with
the fact that OH molecules are effectively produced in the outer parts
of circumstellar envelopes due to photoionization of H$_2$O by
interstellar UV photons. The typical abundance for OH molecules
relative to H$_2$ is about $10^{-5}$ and HST observations
\citep[see][]{sah98} show a compact bright bipolar source expanding
along an axis roughly orthogonal to the bipolar axis.  Despite the fact
that almost all planetary nebulae appear optically thin at 5GHz, Vy 2-2
is optically thick \citep[see][]{pur82}.

\section{Implications on Pumping Schemes}

There exist two main ways to invert the 5cm maser transition, (1) by
radiative pumping of far infrared (FIR) photons or (2) by collisions
with H$_2$ molecules (in combination with local and non-local line
overlap). Chemical pumping does not seem applicable and interstellar UV
photons are only responsible for the dissociation of H$_2$O molecules
in the outer part of the circumstellar envelope to produce OH (the
contribution of stellar UV photons is negligible except perhaps for the
innermost= regions of the envelope).

\subsection{About the ground-state masers}
Theoretical studies of the pumping mechanism of the 18cm OH lines are
quite advanced. The absorption of FIR photons at 34.6 \micron m and
53.3 \micron m excites the OH from the ground state to the \dpi1
ladder.  Subsequent cascading of the populations through the J=1/2 and
J=3/2 levels inverts the J=3/2 ground state \citep{eli76}. This scheme
explains rather well the strong 1612 MHz line and essentially avoids
the \dpi3 ladder \citep{dav79}. Therefore, in circumstellar regions
where 1612 MHz is strong and the above mechanism prevails one should
not expect to detect 5\,cm OH emission.

However, the pump cycle may differ in inner layers where highly excited
lines are expected to be found. Moreover, \cite {col94,col95} argue
that FIR line overlaps are also important to enhance the 1612 MHz line
and may even be the primary inverting scheme for the 1612 MHz line in
not too optically thick OH envelopes. In the latter case overlaps at
120 \micron m are important and populating the J=5/2 level is
essential. \cite{col94} argue that FIR line overlaps alone cannot
explain the main-line masers in stars (contrary to earlier models) and
that near infrared (NIR) overlap effects (with OH or H$_2$O) are likely
needed to explain the main line emission from thin circumstellar
shells.

The most recent model published on OH masers in circumstellar envelopes
\citep[see][]{tha98} treats only the ground-state excitation and
considers two models. The first one is with line overlapping limited by a
Doppler shift of 2km\,s$^{-1}$ and the other one with large overlapping
(up to the expansion velocity). In both models the 1612MHz appears much
stronger than the other ground state lines by a factor 10$^2$ to
10$^3$.  They found that the pumping based on FIR hyperfine line
overlapping is much smaller in the second case.  They suggest that FIR
line overlapping occurs inside clumps (small Doppler shift) of
circumstellar envelopes (this idea was also invoked by Collison and
Nedoluha),
but no prediction is made about the excited state.
Only the recent work of \cite {pav96, pav00} studied excited OH maser
emission but only in the case of massive star forming regions.  

Modeling the detected maser emission in Vy 2--2 at 1612 and 6035 MHz is
a challenge. The particular nature of the source, a very young proto
planetary nebulae, may certainly be a clue. The fact that two masers
are observed at the same LSR velocity (--62 km\,s$^{-1}$) argues in
favor of their spatial association. In such a case they would both
originate from the thin ionization shell presenting similar conditions
as in HII regions.  Within this context, PPN shells may be
characterized by particularly high densities and long path lengths for
coherent amplification that have to be taken into account.

\subsection{Infrared pumping ?}

  Excitation of the OH radical results from complex competitive schemes
involving both collisional and radiative pumping as well as line
overlap effects that are correlated with the velocity field in the OH
medium and local line broadening. The 5 cm OH lines arise from energy
levels 84 cm$^{ -1}$ above the ground-state and we therefore expect
that FIR photons around 100 \micron m are involved in the OH pumping
cycle.  To evaluate the possibility of a pumping scheme based only on
IR photons, Fig.  \ref{flux_iras} compares the IRAS flux at 60 \micron
m and the lower limits of OH emission at 6GHz, assuming that the ratio
between the radio solid angle and the IR solid angle is $\approx 1$.
Fig. \ref{flux_iras} shows that the number of FIR photons largely
exceeds the emitted radio photons.  \cite{bau97} reached a similar
conclusion for compact HII regions but in that case many 5 cm OH masers
could be detected. From this we conclude that in stellar envelopes the
OH pumping mechanism is different from that in massive star forming
regions and that the available FIR radiation is unlikely to
work as a pump for this maser. Moreover as the envelopes are dense, it
is possible that even if the IR pumping were efficient, collisions
could effectively quench the OH maser emission.

\section{Conclusion}

We have observed an extensive sample of OH/IR stars and late-type
variables. Except for one atypical source (Vy 2--2) no excited OH maser
has been detected. We do not confirm the tentative detection of NML
Cyg.  Only the blue shifted emission is detected in Vy 2--2 for which
we have discussed briefly the morphology and OH production. The absence
of detectable excited emission at 5cm (except the unique object Vy
2--2) tends to argue in favor of a pumping scheme based essentially on
the absorption of 35 and 53 $\mu$m photons.  In the case of Vy 2--2 the
ionization shell from where the maser emission seems to originate, may
present physical conditions (shock, higher temperature and density)
similar to those prevailing in HII regions. A hybrid pumping model
applying to both OH/IR stars and HII regions or even a pumping scheme
similar to OH maser emission in massive star forming regions may be
successful. It is interesting to note that no absorption at 35 and 53
$\mu$m can be seen for this source in ISO observations.
Interferometric observations are needed to spatially determine the
likely related positions of the 1612 and 6035 MHz maser emission in Vy
2-2.


\begin{acknowledgements}
  This research has made use of the SIMBAD database operated at CDS,
Strasbourg, France and ASTRID database, operated at GRAAL, Montpellier,
France. We thank Dr A.M.S Richard for valuable suggestions to prepare
the observations and N.J. Rodr\'{\i}guez-Fern\'andez for his help on
ISO data.

\end{acknowledgements}



\begin{thebibliography}{}
  \bibitem[Acker (1978)]{ack78}
          Acker, A., 1978, A\&AS 33, 367.

  \bibitem[Baudry(1974)]{bau74}
           Baudry\,A., 1974, A\&A 32, 191.

  \bibitem[Baudry et al.(1997)]{bau97}
           Baudry\,A., Desmurs\,J.-F., Wilson\,T.L. \& Cohen\,R.J.,
           1997, A\&A 325, 255-268.

  \bibitem[Bensby \& Lundstr\"om(2001)]{ben01}
           Bensby, T. \& Lundstr\"om, I., 2001, A\&A 374, 599. 

  \bibitem[Cesaroni \& Walmsley(1991)]{ces91}
           Cesaroni\,R. \& Walmsley\,M.C. 1991, A\&A 241, 537.

  \bibitem[Christianto \& Seaquist(1998)]{chr98}
           Christianto, H. \& Seaquist, E.R., 1998, AJ 115, 2466.

  \bibitem[Claussen \& Fix(1981)]{cla81}
           Claussen, M.J. \& Fix, J.D. 1981, ApJ 250, L77.

  \bibitem[Clegg \& Walsh(1989)]{cle89}
           Clegg, R. \& Walsh, J.R., 1989, IAU 131, p443.

  \bibitem[Cohen \& Barlow(1974)]{coh74}
           Cohen M. \& Barlow M.J., 1974, ApJ 193,401.

  \bibitem[Cohen et al.(1991)]{coh91}
           Cohen M., Masheder, M.R.W. \& Walker, R.N.F., 1991, 
           MNRAS 250, 611.

  \bibitem[Collison \& Nedoluha(1994)]{col94}
           Collison A.J. \& Nedoluha G.E., 1994, ApJ 422, 193.

  \bibitem[Collison \& Nedoluha(1995)]{col95}
           Collison A.J. \& Nedoluha G.E., 1995, ApJ 442, 311.

  \bibitem[David et al.(1993)]{dav93}
           David P., Le Squeren A.M., Sivagnanam P., 1993, A\&A 277, 453

  \bibitem[Davis et al.(1979)]{dav79}
           Davis, L.E., Seaquist, E.R. \& Purton, C.R.,
           1979, ApJ 230, 434.

  \bibitem[Elitzur et al.(1976)]{eli76}
           Elitzur M., Goldreich, P. \& Scoville, N., 1976, ApJ 205, 384.

  \bibitem[Engels(1979)]{eng79}
          Engels, D., 1979, A\&A sup. 36, 337-345.

  \bibitem[Epchtein et al.(1980)]{epc80}
           Epchtein, N.; Guibert, J., Nguyen-Quang-Rieu, Mr. Turon,
           P. \& Wamsteker, W., 1980, A\&A 85, L1.

  \bibitem[Gray et al.(1992)]{gra92}
           Gray\,M.D., Field\,D. \& Doel\,R.C., 1992, A\&A  262, 555.

  \bibitem[Jewell et al.(1985)]{jew85}
           Jewell, P.R., Schenewerk, M.S. \& Snyder, L. E., 1985, 
           ApJ 295, 183.

  \bibitem[Knapp \& Morris(1985)]{kna85}
          Knapp, G.R. \& Morris, M. 1985, ApJ 292, 640.

  \bibitem[Miranda \& Solf(1991)]{mir91}
           Miranda, L.F.\& Solf, J., 1991, A\&A 252, 331.

  \bibitem[Moore et al.(1988)]{moo88}
           Moore\,T.J.T., Cohen\,R.J., Mountain\,C.M., 1988, MNRAS 231, 887.

  \bibitem[Ott et al.(1994)]{ott94}
           Ott\,M., Witzel\,A., Quirrenbach\,A. et al., 1994, A\&A  284, 331.

  \bibitem[Pavlakis \& Kylafis(1996)]{pav96}
           Pavlakis\,K.G. \& Kylafis\,N.D., 1996, ApJ  467, 300 and 467, 309.

  \bibitem[Pavlakis \& Kylafis(2000)]{pav00}
           Pavlakis\,K.G. \& Kylafis\,N.D., 2000, ApJ 534, 770-780.

  \bibitem[Purton et  al.(1982)]{pur82}
           Purton, C.R., Feldman, P.A., Marsh, K.A., Allen, D.A. \&
           Wright, A.E., 1982, MNRAS 198, 321.

  \bibitem[Sahai \& Trauger(1998)]{sah98}
           Sahai \& Trauger, 1998, ApJ 116, 1357.

  \bibitem[Seaquist \& Davis(1983)]{sea83}
           Seaquist, E.R. \& Davis, L.E., 1983, ApJ 274, 659.

  \bibitem[Sivagnanam et al.(1988)]{siva88}
           Sivagnanam P., Le Squeren A.M., Foy F., 1988, A\&A 206, 285.

  \bibitem[Sylvester et al.(1997)]{syl97} 
           Sylvester, R.J., Barlow, M.J., Liu, X.W., Rieu, Nguyen Q.
           Bach, Truong, Skinner, C.J., Cohen, R.J., Lim, T., Cox,
           P. \& Smith, H.A., 1997, MNRAS 291, L42.

  \bibitem[Thacker et al.(1970)]{tha70}
           Thacker, D.L., Wilson, W.J. \& Barrett, A.H., 1970,
           ApJ 161, L191.

  \bibitem[Thai-Q-Tung et al.(1998)]{tha98}
           Thai-Q-Tung, Dinh-V-Trung, Nguyen-Q-Rieu, Bujarrabal\,V., Le
           Bertre\,T., \& G\'erard\,E, 1998, A\&A 331, 317-327.

  \bibitem[Yen et al.(1969)]{yen69}
           Yen\,J.L., Zuckerman\,B., Palmer\,P., Penfield\,H,  
           1969, ApJ 156, L27.

  \bibitem[Zijlstra et al.(1989)]{zij89}
           Zijlstra, A.A., Te Lintel Hekkert, P., Pottasch, S.R., 
           Caswell, J.L., Ratag, M., Habing, H.J., 
           1989, A\&A 217, 157.

  \bibitem[Zuckerman et al.(1968)]{zuc68}
           Zuckerman\,B., Palmer\,P.,  Penfield\,H., Lilley\,A.E., 
           1968, ApJ 153, L69.

  \bibitem[Zuckerman et al.(1972)]{zuc72}
           Zuckerman\,B., Yen, J.L., Gottlieb, C.A., Palmer, P
           1972, ApJ 177, 59.


\end{thebibliography}
\end{document}